\documentclass[reprint,twocolumn,superscriptaddress,showpacs,nofootinbib,notitlepage,secnumarabic,preprintnumbers,amssymb, nobibnotes, aps, prd, nolongbibliography]{revtex4-2}
\usepackage{graphicx}
\usepackage{latexsym,amsmath,amssymb,lmodern,float,url}
\usepackage{natbib}
\usepackage{color}
\usepackage{microtype}
\usepackage{slashed}
\usepackage{multirow}
\usepackage{comment}
\usepackage{makecell}
\usepackage{tikz}
\usetikzlibrary{arrows,positioning,arrows.meta} 
\tikzset{
    >=stealth',
    punkt/.style={
           rectangle,
           rounded corners,
           draw=black, very thick,
           text width=6.5em,
           minimum height=2em,
           text centered},
    pil/.style={
           ->,
           thick,
           shorten <=2pt,
           shorten >=2pt,}
}
\newcommand{\eq}[1]{Eq.~(\ref{eq:#1})}

\newcommand{\Tr}{\mbox{Tr}}
\newcommand{\re}{\mbox{Re}}
\newcommand{\im}{\mbox{Im}}

\usepackage[colorlinks=true,backref=false, linktocpage=true,
citecolor=blue,urlcolor=blue,linkcolor=blue,pdfpagemode=UseOutlines]{hyperref}

\hypersetup{%
  bookmarksnumbered=true,
  pdftitle = {},
  pdfsubject = {},
  pdfauthor = {},
  pdfkeywords = {}
}

\newcommand{\todo}{\colorbox{red}{\textsc{Todo}}}

\begin{document}
\preprint{FERMILAB-PUB-21-091-T}
\title{Quantum algorithms for transport coefficients in gauge theories}
\author{Thomas D. Cohen}
\email{cohen@umd.edu}
\affiliation{Department of Physics, University of Maryland, College Park, MD 20742, USA}
\author{Henry Lamm}
\email{hlamm@fnal.gov}
\affiliation{Fermi National Accelerator Laboratory, Batavia,  Illinois, 60510, USA}
\author{Scott Lawrence}
\email{scott.lawrence-1@colorado.edu}
\affiliation{Department of Physics, University of Colorado, Boulder, CO 80309, USA}
\author{Yukari Yamauchi}
\email{yyukari@umd.edu}
\affiliation{Department of Physics, University of Maryland, College Park, MD 20742, USA}
\collaboration{NuQS Collaboration}
\date{\today}


\begin{abstract}
In the future, \emph{ab initio} quantum simulations of heavy ion collisions may become possible with large-scale fault-tolerant quantum computers. We propose a quantum algorithm for studying these collisions by looking at a class of observables requiring dramatically smaller volumes: transport coefficients.  These form nonperturbative inputs into theoretical models of heavy ions; thus, their calculation reduces theoretical uncertainties without the need for a full-scale simulation of the collision.  We derive the necessary lattice operators in the Hamiltonian formulation and describe how to obtain them on quantum computers.  Additionally, we discuss ways to efficiently prepare the relevant thermal state of a gauge theory. 
\end{abstract}

\maketitle

\section{Introduction}\label{sec:intro}
The ultimate promise of quantum computers is that physical properties effectively uncalculable by classical algorithms can be obtained~\cite{Feynman:1981tf}. Within high-energy physics, it is believed that quantum algorithms will shed light upon topics where nonperturbative nonequilibrium dynamics play a role~\cite{pqa_loi}. The limitations of current classical methods are particularly acute in heavy-ion collisions.

The standard picture of heavy-ion collisions is that two nuclei collide at high energy, and through multiple scatterings form an expanding fireball of quark-gluon plasma (QGP).  As the plasma cools, the quarks and gluons rehadronize into mesons and baryons that are then measured in the detector. The current theoretical framework divides the collision into overlapping steps which can each be approximated by a semiclassical model~\cite{Gelis:2021zmx}. In particular, the incredible effectiveness of hydrodynamics to model the QGP is notable~\cite{Romatschke:2017ejr,Shen:2020mgh}. While these calculations can account for some of the nonperturbative and nonequilibrium dynamics of QCD at the LHC and RHIC~\cite{Stephanov:1998dy,Stephanov:1999zu,Auvinen:2013sba,Ryu:2015vwa,Alba:2017hhe,Denicol:2018wdp,Dore:2020jye,NunesdaSilva:2020bfs,Reichert:2020oes,Parotto:2018pwx}, the large number of free parameters and uncontrolled systematics used in theoretical modeling are unpalatable~\cite{Cheng:2021tnq,Rocha:2021zcw,Plumberg:2021bme,McLaughlin:2021dph}. Of interest to this work are the transport coefficients: diffusivity~\cite{PhysRevD.70.056001,PhysRevD.94.034025,PhysRevD.96.014032}, conductivity~\cite{PhysRevD.94.034025,PhysRevD.96.014032,PhysRevLett.110.182301}, and viscosity~\cite{PhysRevC.86.024913,PhysRevC.98.014902,PhysRevLett.116.212301,PhysRevC.93.024907,Paquet:2017mny,PhysRevC.98.054906,PhysRevC.101.044904,PhysRevC.101.044904,McLaughlin:2021dph}.

In principle, a complete nonperturbative calculation would be possible from lattice field theory (LFT). Unfortunately, all known formulations of LFT suitable for classical computations use the Euclidean metric in order to allow stochastic methods to sample the path integral. While this allows for a broad range of observables to be computed, its capabilities at finite-density or for Minkowksi observables are practically limited by \emph{sign problems}~\cite{deForcrand:2010ys} or the ill-posedness of analytic continuation~\cite{Tripolt:2018xeo}. Despite these issues, efforts have been undertaken to extract nonperturbative inputs~\cite{PhysRevC.98.054906} --- e.g.~parton distrubtion functions (PDFs)~\cite{Karpie:2019eiq,Cichy:2019ebf,Sufian:2019bol,Alexandrou:2020qtt,Zhang:2020rsx,Huo:2021rpe}, phase diagrams~\cite{Bazavov:2017dus,Borsanyi:2020fev,Giordano:2020roi,Bellwied:2021nrt}, transport coefficients~\cite{Aarts:2002cc,Meyer:2011gj,Kohno:2011aa,Aarts:2014nba,Itou:2020azb,Brambilla:2020siz} --- for the models from lattice QCD, albeit with large uncertainties.

Quantum computers provide a natural facility for studying real-time dynamics of quantum systems. With sufficiently large quantum resources, one should be able to access arbitrary Minkowski matrix elements in LFT; however, such calculations are forbiddingly expensive. For example, consider the task of simulating heavy ion collisions~\cite{deJong:2020tvx}.  This corresponds to an extension of the scattering calculations considered in~\cite{Jordan:2011ci,Jordan:2014tma,Jordan:2017lea}. To prepare well separated wave-packets for the ions, the spatial lattice must be many times the diameter of a heavy nucleus, which is $\sim 10\;\mathrm{fm}$.  To simulate the internal dynamics of each nucleus, a resolution smaller than the nuclear radius is required, thus the lattice spacing should be $\sim 0.1\;\mathrm{fm}$. Together, these two scales imply $\gtrsim 10^6$ lattice sites are required, and so are at least that many qubits. To estimate the circuit depth, one should evolve the calculation for sufficient time that the two nuclei can collide and the final states become sufficiently separated, again $\sim 10\;\mathrm{fm}$~\cite{Trainor:2013bma}. In order to keep the trotterization errors small, one must use a small time step, again $\sim 0.1\;\mathrm{fm}$. This suggests $\sim 10^2$ time steps, ignoring the circuits required for state preparation and measurement.  These estimates are certainly beyond near-term prospects. From this example, we can see that the need to represent multiple scales accurately drives the large resource requirements.

The authors of~\cite{Lamm:2019uyc} pointed out that by decomposing scattering simulations into a nonpertubative input (in that case the hadronic tensor) convolved with perturbative expressions, one reduces the resource requirements.  In that example, instead of preparing two well-separated protons, one prepares a single proton in a $L\sim 1\;\mathrm{fm}$ box.  This reduces qubit costs by a factor of $\sim10^3$.

Heavy-ion collisions allow for a similar decomposition. In particular, the transport coefficients of the QGP should require $L\sim 1\;\mathrm{fm}$. Since the transport coefficients represent the hydrodynamics of the theory, they should require reduced resolution compared to partonic observables.  Using thermodynamic observables of QCD as a guide~\cite{Sharma:2013hsa}, one might anticipate a required lattice spacing (for studying $\sim 200\;\mathrm{MeV}$ matter) of $a \sim 0.1\;\mathrm{fm}$. Further, theoretical models~\cite{Stachel:2013zma,Schlichting:2019abc} and experimental data~\cite{Abelev:2013vea,Abelev:2013xaa,ABELEV:2013zaa} suggest that thermalization happens rapidly -- on the scale of $~1\;\mathrm{fm}$ -- which would reduce the circuit depth. One must further emphasize that the current theoretical uncertainties on the transport coefficients of the QGP plasma are $O(1)$.  Together, these arguments strongly suggest that the transport coefficients of gauge theories represent serious targets for practical quantum advantage in particle physics~\cite{pqa_loi}.

There is another point that weights heavily in favor of focusing upon transport coefficients. For many problems, it is not \emph{a priori} obvious that factoring the physical system into separate regimes is valid. In the case of PDFs, for instance, factorization theorems have been proven only in specific kinematic regimes~\cite{Collins:1989gx}.  It is well-known that using PDFs outside of these regimes leads to issues~\cite{Ma:2014oha,Baumgart:2018ntv}. In heavy-ion collisions, the validity of the hydrodynamic approximation has been studied at length (see~\cite{Romatschke:2017ejr} for a review) and the transport coefficients can be nonperturbatively defined on the lattice~\cite{Meyer:2011gj}.

 Past work has extensively discussed real-time evolution in a gauge theory via digital quantum simulation. Here, we detail novel aspects of quantum simulation in the hydrodynamic regime of gauge theories. In particular, we construct implementations of the stress-energy tensor in the Hamiltonian formulation, allowing hydrodynamic correlators to be measured.

The most involved part of a quantum simulation is typically the preparation of the initial state~\cite{2010PhRvL.105q0405B,PhysRevLett.108.080402,Lamm:2018siq,brandao2019finite,Clemente:2020lpr,Harmalkar:2020mpd,Gustafson:2020yfe,motta2020determining}. For transport coefficients, the desired initial state is in thermal equilibrium. When studying Yang-Mills or QCD, the most interesting temperatures are $\sim 200\;\mathrm{MeV}$. The fact that we are interested in relatively high-temperature states is a key advantage to studying QGP transport over lower temperature processes like scattering: thermal states are markedly easier to prepare. In addition to descriptions of appropriate lattice measurements, this paper details several viable methods for thermal state preparation.

This paper is organized as follows. Sec.~\ref{sec:transport} is devoted to describing different methods of extracting transport coefficients from a quantum simulation --- these approaches are largely independent of the simulation scheme and the construction of lattice observables.  Since hydrodynamic transport coefficients are derived from correlators of the energy-momentum tensor $T_{\mu\nu}$ (EMT), in Sec.~\ref{sec:energy-momentum} we derive lattice operators for the energy-momentum tensor within the Hamiltonian formalism.  Methods for preparing quantum thermal states are elucidated in Sec.~\ref{sec:stateprep}. We conclude in Sec.~\ref{sec:discussion} with some general points and discuss where future theoretical work is required.

\section{Estimating Transport}\label{sec:transport}
Dissipative processes cannot be seen in perfect thermal equilibrium, and so are not accessible to Euclidean lattice calculations. Near equilibrium (and in the limit of small gradients), these processes are characterized by a small number of low-energy constants, termed transport coefficients. The rate of diffusion of a conserved quantity $\varphi$, for instance, is governed by \emph{Fick's law}:
\begin{equation}
    \vec J = -D \frac{d\varphi}{d \vec x}
\end{equation}
where $J$ is the diffusive flux, and $D$, the diffusion constant, is the low-energy constant of interest. In this section, we will describe how to determine such transport coefficients in a quantum simulation.

Many other transport coefficients can be defined, typically related to the behavior of locally conserved quantities. In the case of hydrodynamics, the most interesting are the shear and bulk viscosities $\eta$ and $\zeta$, which are defined by their role in the Navier-Stokes equations:
\begin{widetext}
\begin{equation}\label{eq:ns}
    \rho \left(\frac{\partial u_i}{\partial t} + u_j \nabla_j u_i\right) + \nabla p
    =
    \eta \partial_j \left(\partial_i u_j + \partial_j u_i\right)
    +
    \left(\zeta- \frac 2 3 \eta\right) \partial_i \partial_j u_j
    \text,
\end{equation}
\end{widetext}
where the mass density $\rho$ and velocity $u$ are functions of both space and time, and the pressure $p$ is, in equilibrium, determined from the energy density by the equation of state.  Strictly speaking, in the case of field theory we should use the relativistic Navier-Stokes equations~\cite{Romatschke:2017ejr}, but up to a substitution of the energy density $\epsilon$ for the mass density $\rho$, the nonrelativistic Eq.~(\ref{eq:ns}) yields the same results with easier intuition. In actual heavy ion simulations, it is necessary to include higher-order terms in the gradient expansion, introducing new transport coefficients~\cite{Muller:1967zza,Israel:1976tn,Hiscock:1985zz}.  We do not discuss the determination of these second-order coefficients here.

For these transport coefficients, more or less the same set of methods are available for their computation. For concreteness, in this section we will focus on the shear viscosity. The methods described below generalize easily to bulk viscosity, diffusion, and other transport properties.

Several well-established methods exist for computing the shear viscosity $\eta$ of fluids in molecular dynamics simulations~\cite{hess2002determining}. Three are worth summarizing here; of these, two have natural analogues in quantum simulation of gauge theories.

In the periodic perturbation (PP) method~\cite{hess2002determining,pousaneh2020shear}, one imposes a shearing force $\mathcal F$ on the system. The shear viscosity acts to resist this force, so that in equilibrium, a small shear wave is created. In the limit of small $\mathcal{F}$ and small wavenumbers $k$, the equilibrium ``displacement'' is
\begin{equation}\label{eq:pp}
\mathcal V = \mathcal F \frac{\rho}{\eta k^2}
\text.
\end{equation}
This method works well in molecular dynamics simulations. Unfortunately, it intrinsically involves the imposition of a non-conservative force. Such a force, by definition, cannot be imposed by the addition of any term to the Hamiltonian; equivalently, any simulation of this system must be nonunitary. This renders it inapplicable for calculations of quantum systems.

The transverse current autocorrelation function (TCAF) method~\cite{hess2002determining,pousaneh2020shear} proceeds from the observation that a shear wave, once created, decays exponentially with a decay constant proportional to $\eta$. To see this, consider the linearization of Eq.~(\ref{eq:ns}) about the equilibrium solution $\rho = \rho_0$, $u_i = 0$. The mode $\mathcal U \cos( k x)$ obeys
\begin{equation}\label{eq:ns-linear}
    \rho_0 \frac{\partial \mathcal U}{\partial t} = - \eta
    \left(\vec k \cdot \vec k \mathcal U + \vec k (\vec k \cdot \mathcal U)\right)
    -
    \left(\zeta- \frac 2 3 \eta\right) \vec k (\vec k\cdot \mathcal U)
    \text.
\end{equation}
Thus a shear wave $u_1(\vec x) \propto \cos(k x_2)$ is seen to decay as $e^{-\frac{\eta}{\rho k^2}t}$. Fitting this exponential decay gives the shear viscosity. Conveniently, the fact that the shear viscosity is encoded in the decay constant means that the normalization of the $T_{\mu\nu}$ need not be correct. (In fact, any operator that couples to the appropriate shear mode is likely to show the correct exponential decay.)

In principle, then, we imagine creating a small shear wave in an equilibrated fluid and watching it decay. This requires carefully taking the limit of a small perturbative force. Instead, we can allow ``Maxwell's angels'', \emph{i.e.}~thermodynamic and quantum fluctuations, to create the shear wave. The shear viscosity is now provided by fitting the TCAF at long\footnote{We assume the quantum fluid is described at long wavelengths and times by pure Navier-Stokes, and therefore neglect the phenomenon of long-time tails~\cite{alder1970decay,kovtun2003hydrodynamic}. For realistic fluids, the exponential decay will eventually be cut off by a power-law. This power-law decay can be extracted just as $\eta$ can.} times:
\begin{align}\label{eq:tcaf}
    C(t) =& \int dx\,dy\, \sin(k y) \sin(k x) \langle T_{01}(x,t) T_{01}(y,0)\rangle\nonumber\\&\propto
    e^{-\frac{\eta}{\rho k^2}t}
    \text.
\end{align}
In the context of a classical molecular dynamics simulation, the calculation of $C(t)$ is straightforward: $\int dx\, \sin(k x) T_{01}(x)$ is measured at various times, and the autocorrelations appear in the resulting stochastic time-series. At first blush, the appropriate procedure on a quantum computer is rather different. In particular, because measurements cannot be performed without altering the wavefunction, one might expect to be required to perform one set of measurements for each time at which one wants to measure $C(t)$. This is an expensive prospect.

Happily, we are explicitly interested in the behavior of \emph{hydrodynamic} fluctuations, whose apparent behavior is classical. The measurement of the amplitude of the shear wave contains very little information relative to the whole wavefunction. This measurement therefore constitutes a sort of ``weak measurement,'' which can be made without unduly disturbing the quantum state\footnote{The notion of a weak masurement --- see e.g.~\cite{Clerk:2008tlb} --- is here simply used to denote a measurement from which very little information is obtained about the state.}.  Therefore, we can measure the amplitude of the shear wave at one time, continue evolving, and subsequently measure the amplitude at many later times.

Similarly, the energy density may be measured in the simulated system prior to any time evolution is done, without significantly disturbing the wavefunction. In fact, as long as the simulated system can be kept coherent, the TCAF can be determined to arbitrary precision with one long time evolution --- in the large-volume limit. Away from the large-volume limit, the number of measurements that can be made can be heuristically expected to be linear in the volume of the system.

The third method follows from the Green-Kubo relation~\cite{zwanzig1965time} connecting shear viscosity to the $\omega,k\rightarrow 0$ limit of the two-point correlator of $T_{12}$:
\begin{equation}\label{eq:gk}
\eta = \frac{V}{T} \int_0^\infty \langle T_{12}(t) T_{12}(0)\rangle
\text.
\end{equation}
This method is reported to converge slowly in molecular dynamics simulations~\cite{hess2002determining}, but has nevertheless proven useful in practice, \emph{e.g.}~\cite{fernandez2004molecular}.
Note that this method, unlike the TCAF method, does require the operator for the stress tensor to be correct, including the normalization.

These methods for computing $\eta$ have natural analogs for most other transport coefficients. However, in the case of bulk viscosity, which is the next-most relevant hydrodynamic property, there is no simple analog of the TCAF method. Instead, molecular dynamics calculations of the $\zeta$ typically proceed from the Green-Kubo relation, which is in terms of the diagonal components of the $T_{\mu\nu}$~\cite{jaeger2018bulk}:
\begin{equation}\label{eq:bulk-gk}
\zeta 
=
\frac{V}{9T} \int_0^\infty d t\; \langle T_{ii}(t) T_{jj}(0)\rangle
\text.
\end{equation}

\section{Energy-Momentum Tensor}\label{sec:energy-momentum}
Transport coefficients in quantum field theory are defined in terms of the $n$-point correlators of $T_{\mu\nu}$. Thus, in order to extract them from lattice calculations, one must define a lattice $T_{\mu\nu}$. Here, we derive such operators in the lattice Hamiltonian formalism natural for use in quantum simulations. In Sec.~\ref{sec:emtonlattice}, we summarize the discretization of the EMT on a spacetime lattice (as distinct from the spatial-only Hamiltonian lattice). In Sec.~\ref{sec:transfer} we review the derivation of Kogut-Susskind Hamiltonian from the transfer matrix~\cite{Creutz:1976ch} which provides a framework for deriving the Hamiltonian formulation of the EMT. In the rest of the section, we derive the necessary components of $T_{\mu\nu}$ for quantum simulation: diagonal components in \ref{sec:Tmumu}, spatial components $T_{ij}$ in \ref{sec:Tij}, and time-like components $T_{0i}$ in \ref{sec:T0i}.

We provide both naive operators (which have $O(a)$ corrections) and tree-level improved operators analogous to the `clover' of the spacetime lattice (these have $O(a^2)$ corrections). The operators derived are summarized in Table~\ref{tab:ops}. They are constructed from link operators $\hat U$ and their conjugate momenta $\hat \pi$, defined in Sec.~\ref{sec:transfer} below. The plaquette $\hat P$ and clover $\hat C$ are defined in Sec.~\ref{sec:emtonlattice}.

\begin{table*}[t]
\caption{Gauge-invariant lattice operators in the Hamiltonian formalism in $3+1d$ dimensions: naive operators with $O(a)$ errors and improved operators with errors that are $O(a^2)$. Components of the energy-momentum tensor $T_{\mu\nu}$ are constructed as linear combinations of these operators according to \eq{em}.\label{tab:ops}  The plaquette $\hat P$ and clover $\hat C$ are defined in Eq.~(\ref{eq:plaq}) and Eq.~(\ref{eq:clover}), respectively. Spatial indices are $i\neq j\neq k$.}
    \centering
    \begingroup
  \renewcommand{\arraystretch}{2.5}    
    \begin{tabular}{c|c|c}
    \hline\hline
        Operator & $O(a)$ & $O(a^2)$ \\
         \hline
         $\Tr F_{0i} F_{0i}(n)$ & $\frac{g_s^2}{a^4} \Tr\left[\pi_{n,i}^2  \right]$& $\sum_{{x=0,1}}\frac{g_s^2}{2a^4} \Tr\left[\pi_{n-x\hat i,i}^2  \right]$\\[.5em]
         $\Tr F_{0i} F_{0j}(n)$ & $\frac{g_s^2}{a^4} \Tr\left[\pi_{n,i} \pi_{n,j}  \right]$ &\makecell{$\frac{g_s^2}{4a^4} \left(
 \Tr\left[\hat\pi_{n,i}\hat\pi_{n,j}  \right] +  \Tr\left[\hat\pi_{n,i}\hat U_{n - \hat j,j}^{\dagger}\hat\pi_{n - \hat j,j}\hat U_{n - \hat j,j} \right]+ \Tr\left[\hat U_{n - \hat i,i}^{\dagger}\hat\pi_{n - \hat i}\hat U_{n - \hat i,i}\hat\pi_{n,j} \right]\right.$\\[.5em]
 $\left.
+\Tr\left[\hat U_{n - \hat i,i}^{\dagger}\hat\pi_{n - \hat i,i}\hat U_{n - \hat i,i}\hat U_{n - \hat j,j}^{\dagger}\hat\pi_{n - \hat j,j}\hat U_{n - \hat j,j} \right] \right)$}\\
         $\Tr F_{0j} F_{ij}(n)$ & $- \frac{1}{a^4}\Tr\left[ \hat \pi_{n,j} \im \hat P_{ij}(n) \right]$ & $-\frac{1}{2a^4}\left( \Tr\left[ \hat \pi_{n,j} \im \hat C_{ij}(n) \right]
    + \Tr\left[ \hat U_{n-\hat j,j}^{\dagger}\hat \pi_{n-\hat j,j} \hat U_{n-\hat j,j}\im \hat C_{ij}(n) \right]\right)$\\
         $\Tr F_{ij} F_{ij}(n)$ & $\frac{2}{g_s^2a^4}\re\Tr\left[1-\hat P_{ij}(n)  \right]$ &$\sum_{x=0,1}\sum_{y=0,1}\frac{1}{2g_s^2a^4}\re\Tr\left[1-\hat P_{ij}(n-x\hat i-y\hat j)  \right]$\\
         $\Tr F_{ij} F_{kj}(n)$ & $\Tr[\hat F_{ij}^N(n) \hat F_{kj}^N(n)]$ & $\Tr[\hat F_{ij}^C(n) \hat F_{kj}^C(n)]$\\\hline\hline
    \end{tabular}
    \endgroup
\end{table*}

\subsection{Energy-momentum tensor on a lattice}\label{sec:emtonlattice}

The energy-momentum tensor is the Noether's current of spacetime translational symmetry. In LFT, this symmetry is explicitly broken to a discrete subgroup and thus naive lattice currents may not be conserved. We expect to restore the symmetry (i.e. the Ward identity) in the continuum limit, albeit renormalization is required. Another tactic instead constructs combinations of the low-dimension operators that mix under the discrete rotations and translations~\cite{Caracciolo:1991cp,Caracciolo:1989pt,Giusti:2015daa} such that a lattice Ward identity is satisfied. A final method extracts a UV-finite version of the EMT from gradient flow~\cite{Suzuki:2013gza,DelDebbio:2013zaa}. In this work, we study two lattice EMTs: the naive EMT accurate to $O(a)$ and the clover EMT with $O(a^2)$ errors. There exists an $O(a^4)$ EMT~\cite{BilsonThompson:2002jk}, but it is left for future work, as it likely first requires developing an appropriately improved Hamiltonian~\cite{Luo:1998dx,Carlsson:2001wp,Luo:1993af}.

\begin{figure}[b]
  \begin{center} 
    \includegraphics[width=0.9\linewidth]{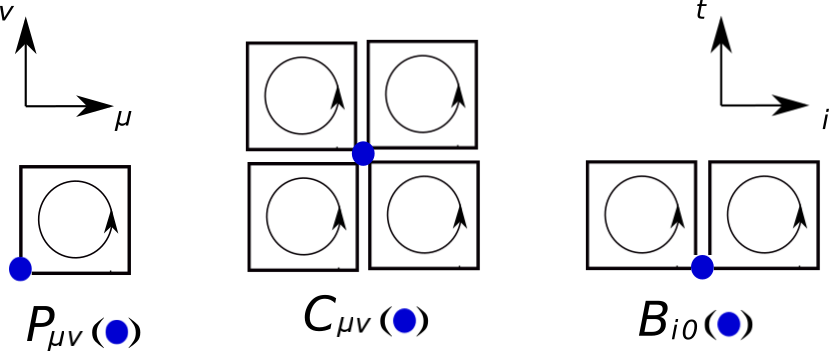}
    \caption{Schematic view of the plaquette, the clover, and the half-clover.} 
    \label{fmunu} 
  \end{center}
\end{figure}

In the continuum, the EMT of a gauge theory is
\begin{equation}\label{eq:em}
    T_{\mu\nu} = \frac{1}{4}g_{\mu\nu}\Tr\left[F_{\alpha\beta}F^{\alpha\beta}\right] - \Tr\left[F_{\mu\alpha}F_{\nu}^{\alpha}\right].
\end{equation}
where the mostly minus convention is used. Following the notation of~\cite{Creutz:1976ch}, we normalize the group generators $\lambda^a$ to
\begin{equation}\label{eq:normG}
\Tr[\lambda^a\lambda^b] = \delta_{ab}
\end{equation}
and functions such as $F_{\mu\nu}$ are defined with this normalization. Before we discretize \eq{em}, we must establish some lattice notation. Links are denoted as $U_{n,\mu}$ where $n$ is the site the link starts at, and $\mu$ is the direction of the link. The fundamental gauge-invariant object, the plaquette, is defined as the product of four links around a closed loop as in Fig.~\ref{fmunu}:
\begin{equation}\label{eq:plaq}
P_{\mu\nu}(n) = U_{n,\mu}U_{n+\hat \mu, \nu}U_{n+\hat \nu, \mu}^{\dagger}U_{n,\nu}^{\dagger}.
\end{equation}

On a spacetime lattice, the terms in the EMT are polynomials of plaquettes. Note that, as terms in \eq{em} contain products of $F_{\mu\nu}$ in different directions, they are most sensibly computed either at lattice sites or at the center of spacetime volumes. For the sake of simplicity, we will focus on discretizing the EMT on lattice sites.

The first term in \eq{em} is the Lagrangian. On the lattice, it can be written
\begin{equation}\label{eq:f2}
\Tr\left[F_{\mu\nu}(n)^2\right] = \frac{2}{g_s^2a_{\mu}^2a_{\nu}^2} \re\Tr[1-P_{\mu\nu}(n)] + O(a)
\text,
\end{equation}
where $g_s$ is the coupling constant. 
Here $a_{\mu}$ are the lattice spacings in the $\mu$ direction. We denote the temporal spacing $a_0$, and assume all other spacings to be the same and denote them as $a$. When working in the Hamiltonian formalism, we take $a_0\rightarrow 0$ before performing continuum extrapolations. If instead we considered $\Tr\left[F_{\mu\nu}(n+1/2[\hat{\mu}+\hat{\nu}])^2\right]$ located at the center of the plaquette, we would find \eq{f2} is accurate to $O(a^2)$. 

To improve the discretization on a site up to $O(a^2)$, we can simply average over the RHS of \eq{f2} for four plaquettes around the site $n$ in $\mu\nu$ plane:
\begin{eqnarray}\label{eq:f2imp1}
    \Tr\left[F_{\mu\nu}(n)^2\right] &=&\sum_{x={0,1}}\sum_{y={0,1}} \frac{1}{2g_s^2a_{\mu}^2a_{\nu}^2}\nonumber\\ \re\Tr[1 &-& P_{\mu\nu}(n-x\hat\mu-y\hat\nu)] + O(a^2) \text.
\end{eqnarray}
In cases of $F_{0i}$, the $O(a_0)$ error induced in the time direction is acceptable as we take the limit $a_0 \ll a$ in the Hamiltonian formalism. This implies that we don't need to average over four plaquettes in both direction $\hat t$ and $\hat i$. Instead, we need to average over only two plaquettes in $\hat i$ direction around the site:
\begin{align}\label{eq:f2imp2}
    \Tr\left[F_{oi}(n)^2\right] =&\sum_{x={0,1}} \frac{1}{g_s^2a_{\mu}^2a_{\nu}^2}\nonumber\\ \re\Tr&[1-P_{\mu\nu}(n-x\hat i)] + O(a^2,a_0)\text.
\end{align}

The second term of \eq{em}, $F_{\mu\alpha}F_{\nu\alpha}$, requires us to construct a discretization for $F_{\mu\nu}$ itself. The naive discretization for the field strength tensor is
\begin{equation}\label{eq:fn}
    F_{\mu\nu}^{N}(n) = -\frac{i}{2g_sa_{\mu}a_{\nu}}\left( P_{\mu\nu}(n) - P_{\mu\nu}^{\dagger}(n) \right) + O(a)\text.
\end{equation}
Note again that $F_{\mu\nu}^N(n)$ is evaluated on a lattice site. The RHS of \eq{fn} approximates the value  at the center of the plaquette up to $O(a^2)$. To improve the on-site discretization scheme up to $O(a^2)$, one can use so-called `clover' operators as shown in Fig.~\ref{fmunu}:
\begin{equation}\label{eq:clover}
C_{\mu\nu}(n) = \frac 1 4 \left[
P_{\mu,\nu}
+
P_{\nu,-\mu}
+
P_{-\mu,-\nu}
+
P_{-\nu,\mu}
\right](n)
    \text.
\end{equation}
From these clover operators we construct the improved clover discretization of the field strength:
\begin{equation}\label{eq:fc}
    F_{\mu\nu}^{C}(n) = -\frac{i}{2g_sa_{\mu}a_{\nu}}\left( C_{\mu\nu}(n) - C_{\mu\nu}^{\dagger}(n) \right) + O(a^2)\text.
\end{equation}
Here we have used $P_{-\mu,\nu}(n)$, for instance, to denote the plaquette in which the first link begins at site $n$ and ends at site $n-\hat\mu$. In an Abelian theory, we of course have $P_{-\mu,\nu}(n) = P_{\mu,\nu}(n-\mu)^\dagger$, as the starting site does not matter. This is not the case in general.

The clover improvement ensures that the leading discretization errors are $O(a^2,a_0^2)$ regardless of orientation. However, as mentioned, working in the Hamiltonian formalism implies $a_0 \ll a$. As a result, $O(a_0)$ corrections are acceptable while $O(a)$ are not. This, combined with the difficulty of implementing time-nonlocal operators, motivates the ``half-clover'' operator --- analogous to \eq{f2imp2} --- averaged over two plaquettes as in Fig.~\ref{fmunu}:
\begin{equation}
    B_{i0}(n) = \frac{1}{2}\left(P_{i0}(n) + P_{0(-i)}(n)\right)\text.
\end{equation}
This is enough to implement $T_{\mu\nu}$ correctly up to $O(a^2,a_0)$:
\begin{equation}\label{eq:fb}
F^{B}_{i0}(n) = \frac{-i}{2g_sa_0a}\left(B_{i0}(n) - B_{i0}^{\dagger}(n)\right).
\end{equation}

\subsection{Transfer matrix}\label{sec:transfer}
With a choice of discretization of $F_{\mu\nu}$ and $F_{\mu\nu}^2$, one can compute a lattice $T_{\mu\nu}$ in the action formulation.
In the Hamiltonian formulation, $T_{\mu\nu}$ (which are functions of the space of field configurations) must be replaced by operators $\hat T_{\mu\nu}$ which act on the Hilbert space of gauge links on a spatial lattice. A few options exist for performing this transformation. One is to use the Legendre transform; however, this is complicated by gauge invariance. The transfer matrix formalism has the advantage of being manifestly gauge-invariant.

Given a path integral, we can construct a Hilbert space and a transfer matrix such that the transfer matrix completely characterizes the path integral. Taking the logarithm of the transfer matrix (in our case, in the $a_0\rightarrow 0$ limit) yields a Hamiltonian usable for quantum simulations.

Now, consider a perturbation to that action by a term proportional to $\mathcal{O}$. The free energy, differentiated with respect to the perturbation, yields the $\langle \mathcal{O}\rangle$. Connecting the perturbed system to one in the Hamiltonian formalism via the transfer matrix, we obtain a perturbed Hamiltonian, revealing what operator $\hat{ \mathcal{O}}$ to use on a quantum computer.

The Hamiltonian which gives the same dynamics as the Wilson action is derived in~\cite{Creutz:1976ch}; this is the unperturbed case of the above procedure. In this section, we first summarise the derivation of the Kogut-Susskind Hamiltonian $H_{KS}$~\cite{PhysRevD.11.395} via the transfer matrix, and then discuss the perturbed case which yields specific operators of interest. Our starting point is the Minkowski path integral corresponding to the Wilson action\footnote{Strictly speaking, the Wilson action is in Euclidean space, with the sign of the second term of Eq.~(\ref{eq:action}) flipped. This section can be formulated in metric with no difference provided the Hamiltonian limit is taken~\cite{Kanwar:2021tkd}; we have chosen Minkowski to preserve a more straightforward correspondence with the quantum simulation.}:
\begin{eqnarray}\label{eq:action}
S_{W} &=& \sum_{t=1}^{N_t} K(t) + V(t)\\
K(t) &=& \sum_{n}\sum_{i}  \frac{a}{g_s^2a_0}\re\Tr\left[1- P_{0i}(n,t)\right]\\
V(t) &=& \sum_{n}\sum_{i<j} \frac{a_0}{g_s^2 a}\re\Tr\left[ P_{ij}(n,t)\right]
\end{eqnarray}
Here $n$ denotes a spatial site and $i,j$ denote spatial directions. The corresponding path integral is of course the integral over all field configurations of $e^{iS}$. Given a Hamiltonian $H$, we could also construct a path integral by splitting the time evolution operator $e^{-i H t}$ into a product of many nearly-identity pieces $\hat T = e^{-i H a_0}$ (this is the transfer matrix) and inserting a complete set of states:
\begin{equation}
    Z =
    \int \mathcal D U
    \langle U_t|\hat P \hat T| U_{t-a_0}\rangle
    \cdots
    \langle U_{a_0} | \hat P \hat T | U_0\rangle
    =
    \int \mathcal D U\;e^{i S}\text.
\end{equation}
Here $|U\rangle$ is a basis state in ``position basis'': an element of the gauge group is specified at each link. The gauge projection operator $P$ has been inserted between every pair of states in order to obtain a time-translation-invariant action.

As a result, we see that for the Hamiltonian $H$ to give the same dynamics as the action of Eq.~(\ref{eq:action}), the matrix elements of the transfer matrix $\hat T=e^{-ia_0\hat H}$ should be given by
\begin{widetext}
\begin{equation}\label{eq:t}
\langle U' | \hat T | U \rangle = e^{i(K(t) + V(t))}
= e^{i\sum_{n} \left(\frac{a}{g_s^2a_0}\sum_{i}\re\Tr\left[1- U_{n,i}U^{'\dagger}_{n,i}\right] + \frac{a_0}{g_s^2a}\sum_{i,j}\re\Tr\left[U_{n,i}U_{n+\hat i,j} U_{n+\hat j,i}^{\dagger}U_{n,j}^{\dagger}\right] \right)}
\text.
\end{equation}
\end{widetext}
We are working in temporal gauge: $U_{n,0} = 1$ for all $n$. For our derivations this is an irrelevant technicality, but see \cite{Creutz:1976ch,Lamm:2019bik} for detailed expositions of the relationship between timelike links and the gauge projection operator.

Eq.~(\ref{eq:t}) suffices to define the transfer matrix, but we would like to express it in terms of more natural objects in the Hamiltonian formulation: in particular, the $\hat U$ that are diagonal in the position basis, and their conjugate operators $\hat \pi$ defined below. As spatial plaquettes are already written only with $U$s, they remain in the same form and all arguments $U$ become operators $\hat U$. On the other hand $K$ results in operators not diagonal in this basis, and so one cannot read off the kinetic part of $H$ from \eq{t} directly. To find an operator which satisfies \eq{t}, we introduce unitary operators defined on each link,
\begin{equation}
\hat R_{n,i}(g)|U_{m,j}\rangle = 
\begin{cases}
|g\, U_{m,j} \rangle \mbox{\,\,\,when\,\,}m=n\mbox{\,\,and\,\,}i=j\\
|U_{m,j} \rangle\mbox{\,\,\,\,\,\,\,otherwise}
\end{cases}
\end{equation}
which can be written with Hermitian operators $\hat\pi$ as
\begin{equation}
    \hat R_{n,i}(g) = e^{i x^a \hat\pi^a_{n,i}}.
\end{equation}
In short, $\hat\pi$ generate rotations $\hat R$ of each link. Here, in the case of $SU(N)$, the $x^a$ are $N^2-1$ real parameters parameterizing the group element $g \in SU(N)$, and $\hat\pi^a_{n,i}$ are eight Hermitian operators (differential operators on the group $SU(N)$) associated to $x^a$ defined on each link. Using these operators, we rewrite the transfer matrix
\begin{align}
\hat T = \int_{g\in G}& \prod_{n,i}\left[dg_{n,i}\;\hat R_{n,i}(g_{n,i})\right]\nonumber\\
&\times e^{i\frac{a}{g_s^2a_0}\sum_{n,i}\re\Tr\left[1- g^{\dagger}_{n,i}\right] + i \hat V}\text.
\end{align}
Exchanging $g$ for real parameters $x^a$, one obtains
\begin{eqnarray}
\hat T = \int_{-\pi}^{\pi} &&\prod_{n,i,a}\left[dx_{n,i}^a\;e^{ix_{n,i}^a\hat\pi_{n,i}^a}\right] \nonumber\\
&\times&e^{i\frac{a}{g_s^2a_0}\sum_{n,i}\re\Tr\left[ 1-e^{ix_{n,i}^a\lambda^a}\right]+ i\hat V}
\text.
\end{eqnarray}
In the limit $a_0\rightarrow 0$, the integral can be evaluated via the saddle point method. The saddle point in this case is degenerate, being the gauge orbit of the point $x=0$. As the whole expression of $\hat T$ is gauge invariant, we take the saddle-point approximation around $x=0$ to obtain
\begin{eqnarray}
\hat T&\sim& \int dx\; e^{i x_{\rho}\hat\pi_{\rho} + \frac{ia}{2 g_s^2a_0}x_{\rho}x_{\rho}+i\hat V}=N e^{-ia_0\frac{g_s^2}{2a}\hat\pi_{\rho}\hat\pi_{\rho}+i\hat V}\text.
\end{eqnarray}
For brevity we have written combined three indices into $\rho \equiv (n,i,a)$. Together with the spatial plaquette terms from $V$, the Hamiltonian is
\begin{equation}
\hat H_{KS} = \frac{g_s^2}{2a}\sum_{n,i,a} \hat\pi_{n,i}^a \hat\pi_{n,i}^a - \frac{1}{g_s^2 a} \sum_{\substack{n\\i<j}} \re\Tr\left[\hat P_{ij}(n)\right]
\text.
\end{equation}
The first term in the Hamiltonian can also be written as $\Tr[\hat\pi^2_{n,i}]$ by defining the $N \times N$ matrix $\hat \pi_{n,i} = \hat \pi^a_{n,i} \lambda^a$. Under gauge transformations, we have
\begin{equation}
    \hat \pi_{n,i} \rightarrow g_n^{-1} \hat \pi_{n,i} g_n\text,
\end{equation}
and this implicitly defines the transformation law for the operators $\hat \pi^a_{n,i}$ as well.
Note that the Hamiltonian in either form is Hermitian and manifestly gauge-invariant.

In the rest of the section, we repeat this procedure in the presence of a perturbation $\mathcal O$ --- in the sections below this perturbation will be $T_{\mu\nu}$. In the action formulation, one way to measure the expectation value of  $\mathcal{O}$ at time $t_0$ in a system governed by the action $S$ is to perturb the action by the function $\mathcal{O}(U)$ and define 
\begin{equation}
    Z_\epsilon = \int \mathcal{D} U \;e^{i(S_0 + \epsilon \mathcal{O}(t_0))}\text.
\end{equation}
Differentiating $Z_\epsilon$ --- or rather its logarithm, the free energy --- with respect to $\epsilon$ yields:
\begin{equation}
    i \langle \mathcal{O}(t_0) \rangle = Z_0^{-1}\left(\frac{\partial Z_\epsilon}{\partial \epsilon} \right)_{\epsilon\rightarrow 0}
    \text.
\end{equation}
A perturbed Hamiltonian $H_\epsilon$, yet to be determined, corresponds to this perturbed path integral. The corresponding transfer matrix has matrix elements given by
\begin{equation}
    \langle U'| e^{-i a_0 H_\epsilon} | U \rangle = e^{i(S + \epsilon \mathcal{O})}\text.
\end{equation}
Differentiating the transfer matrix with respect to $\epsilon$ yields $i$ times the desired operator; therefore we see the Hamiltonian should be perturbed to
\begin{equation}
    \hat H_\epsilon = \hat H_0 - \epsilon \hat O / a_0
    \text.
\end{equation}
So, by perturbing the Lagrangian with a component of $T_{\mu\nu}$ and finding the corresponding perturbed Hamiltonian $\hat H_\epsilon$, one can read off the operators $\hat T_{\mu\nu}$. To be precise, if the perturbing parameter is $\epsilon$, then the corresponding operator in the Hamiltonian formalism is the coefficient of $(\epsilon)$.

\subsection{\texorpdfstring{$T_{\mu\mu}$}{Tmumu} in the Hamiltonian formulation}\label{sec:Tmumu}
The diagonal components of the EMT consist of two kinds of terms: $F_{0i}F_{0i}$ and $F_{ij}F_{ij}$. Without loss of generality, let us consider $T_{11}$ in two spatial dimensions. Perturbing the action by terms proportional to $F_{01}(n_0)^2$ and $F_{12}(n_0)^2$, we have
\begin{equation}
S_\epsilon = S_W + \epsilon a_0 a^3 \Tr\left[F_{01}(n_0)^2+  F_{12}(n_0)^2  \right]
\text.
\end{equation}
To simplify notation, $S_\epsilon$, $H_\epsilon$, and $T_\epsilon$ will have different meanings in this and each subsequent subsection, corresponding to the different types of perturbations being considered.

The perturbing terms are defined on the spacetime lattice via \eq{f2}. For $H_\epsilon$ to give the same dynamics as by the action $S_\epsilon$, they should be connected via transfer matrix $T_\epsilon = e^{-i a_0 H_\epsilon}$ as
\begin{equation}\label{eq:t1}
    \langle U' | T_\epsilon | U\rangle = e^{iS_{\epsilon}} = e^{i (K_{\epsilon}+V_{\epsilon})}
\end{equation}
with 
\begin{eqnarray}
    K_{\epsilon}(U',U) &=& K + \epsilon  \frac{2a}{g_s^2a_0}\re\Tr\left[1- U_{n_0,1}U^{'\dagger}_{n_0,1}\right]\label{eq:t11}\\
    V_{\epsilon}(U',U) &=& V + \epsilon  \frac{2a_0}{g_s^2 a}\re\Tr\left[1-P_{12}(n_0)\right] \label{eq:t12}
\end{eqnarray}

The spatial plaquettes in \eq{t12} correspond to diagonal operators; it is only for \eq{t11} that the transfer matrix formalism is useful. A little algebra verifies that this Lagrangian corresponds to the transfer matrix
\begin{equation}
\hat T_\epsilon = \int\mathcal{D}g
\; e^{iK(g)  + i\epsilon \frac{2a}{g_s^2a_0}\text{ReTr}[1-g_{n_0,1}^{\dagger}] + i \hat V_{\epsilon}}
\end{equation}
where we have introduced shorthand
\begin{eqnarray}
    \int\mathcal{D}g &\equiv& \int_{g\in G} \prod_{n,i} dg_{n,i} \hat R_{n,i}(g_{n,i})\\
    K(g) &\equiv& \frac{a}{g_s^2 a_0}\sum_{n,i}\re\left[1-g^{\dagger}_{n,i} \right].
\end{eqnarray}
Performing the saddle-point approximation gives:
\begin{eqnarray}
\hat T_\epsilon&\sim& \int dx\; e^{i x_{\rho} \hat\pi_{\rho} - x_{\rho} M_{\rho\sigma}  x_{\sigma} + i\hat V_{\epsilon}}\\
&\rm{with\,\,}& M_{\rho\sigma} = \frac{-ia}{2g_s^2a_0}\delta_{\rho\sigma} - \epsilon\frac{ia}{g_s^2a_0}\delta_{nn_0}\delta_{mn_0}\delta_{i1}\delta_{j1}\delta_{ab}\nonumber\text.
\end{eqnarray}
Here we have abbreviated $\rho \equiv (n,i,a)$ and $\sigma\equiv(m,j,b)$. The integral gives to $O(\epsilon)$,
\begin{equation}
\hat H_\epsilon = \hat H_{KS} - \epsilon \left(\frac{g_s^2}{a}\Tr\left[\hat\pi_{n_0,1}^2\right] + \frac{2}{g_s^2a} \re\Tr\left[1-\hat P_{12}(n_0)\right]\right)
\end{equation}
We read off the operators for $F_{\mu\nu}F_{\mu\nu}$:
\begin{eqnarray}
\Tr\left[\hat F_{0i}(n_0)^2\right] &=& \frac{g_s^2}{a^4} \Tr\left[\hat\pi_{n_0,i}^2\right] \\
\Tr\left[\hat F_{ij}(n_0)^2\right] &=& \frac{2}{g_s^2a^4} \re\Tr\left[1-\hat P_{ij}(n_0)\right] 
\text.
\end{eqnarray}
Finally, we can construct $T_{\mu\mu}$ from these operators:
\begin{eqnarray}\label{eq:tmm}
\hat T_{00}(n_0) &=& \frac{g_s^2}{2a^4}\sum_{i}\Tr[\hat\pi_{n_0,i}^2]\nonumber\\ &+&\frac{1}{g_s^2a^4}\sum_{i<j}\re\Tr[1-\hat P_{ij}(n_0)]\\
\hat T_{ii}(n_0)&=& \frac{g_s^2}{2a^4}\Tr\left[-\hat\pi_{n_0,i}^2+\hat\pi_{n_0,j}^2+\hat\pi_{n_0,k}^2  \right]\nonumber\\
+\frac{1}{g_s^2a^4}&\re&\Tr\left[ 1-\hat P_{ij}(n_0)-\hat P_{ik}(n_0)+\hat P_{jk}(n_0)\right]
\end{eqnarray}
Note that $T_{00}(n)$ is the density of the Kogut-Suskind Hamiltonian $H_{KS}$ up to a constant term. Because we have worked only at tree level, the trace of the EMT operator vanishes as is the case in \eq{em}.

Eq.~(\ref{eq:tmm}) gives $\hat T_{\mu\mu}$ up to $O(a)$. To improve the $\hat T_{\mu\mu}$ operators up to $O(a^2)$, we use \eq{f2imp1} and \eq{f2imp2} and take the average around the site $n_0$:
\begin{eqnarray}
\Tr\left[\hat F_{0i}(n_0)^2\right] &=& \sum_{x=0,1}\frac{g_s^2}{2a^4} \Tr\left[\hat\pi_{n_0-x\hat i,i}^2\right] \\
\Tr\left[\hat F_{ij}(n_0)^2\right] &=& \sum_{x=0,1}\sum_{y=0,1}\frac{1}{2g_s^2a^4} \nonumber\\
&&\re\Tr\left[1-\hat P_{ij}(n_0-x\hat i-y\hat j)\right] 
\text.
\end{eqnarray}
These operators enable us to construct $\hat T_{\mu\mu}$ up to discretization errors that are $O(a^2,a_0)$.

\subsection{\texorpdfstring{$T_{ij}$}{Tij} in the Hamiltonian formulation}\label{sec:Tij}
Let us now move to deriving the operators $\hat T_{ij}$, that is, the off-diagonal spatial parts of the EMT:
\begin{equation}
T_{ij} = \Tr\left[-F_{i0}F_{j0} + F_{ik}F_{jk}\right]\label{eq:tij}
\text.
\end{equation}
This definition holds both on the spacetime lattice and as an operator equation on the Hamiltonian lattice. We first work with the naive discretization, and then with the clover discretization.
Without loss of generality, let us take $T_{12}$ as an example and perturb the Wilson action with terms in \eq{tij}. We find that $\hat T_\epsilon$ is given by Eq.~(\ref{eq:t1})
with:
\begin{eqnarray}
    K_{\epsilon} &=& K + \epsilon a_0a^3  \Tr\left[F^{N}_{10}F^{N}_{20}\right]\label{eq:t21}\\
    V_{\epsilon} &=& V + \epsilon a_0a^3  \Tr\left[F^{N}_{13}F^{N}_{23}\right]\label{eq:t22}
    \text.
\end{eqnarray}
As before, the spatial plaquettes in \eq{t22} can be directly converted to operators. The time-like plaquettes in \eq{t21} will ultimately appear as various $\hat\pi$. Using $\hat R(g)$ operators, $\hat T_\epsilon$ can be written
\begin{equation}
\hat T_\epsilon = \int\mathcal{D}g
\; e^{i K(g)-i\epsilon\frac{a}{4g_s^2a_0}\text{Tr}\left[ (g_{n_0,1}^{\dagger}-g_{n_0,1})(g_{n_0,2}^{\dagger}-g_{n_0,2}) \right] + i\hat  V_{\epsilon}}\text.
\end{equation}
Evaluating the integral via the saddle point $x=0$ (exact in the limit $a_0\rightarrow 0$) gives:
\begin{eqnarray}
\hat T_\epsilon&\sim& \int dx\; e^{i x_{\rho} \hat \pi_{\rho} - x_{\rho} M_{\rho\sigma}  x_{\sigma}+ i \hat V_{\epsilon}} = A e^{-\frac{1}{4}\hat \pi_{\rho}M^{-1}_{\rho\sigma}\hat \pi_{\sigma}+ i \hat V_{\epsilon}}\nonumber\\
&& M_{\rho\sigma} = \frac{-ia}{2g_s^2a_0}\delta_{\rho\sigma} + \epsilon\frac{ia}{g_s^4a_0}\delta_{nn_0}\delta_{mn_0}\delta_{i1}\delta_{j2}\delta_{ab}\label{eq:it2}
\end{eqnarray}
which at $O(\epsilon)$ yields $\hat H_\epsilon$:
\begin{equation}
    \hat H_\epsilon = \hat H_{KS} - \epsilon \frac{g_s^2}{a}\Tr[\hat \pi_{n_0,1}\hat \pi_{n_0,2}] -  \epsilon a^3\Tr\left[\hat F^{N}_{ik}(n_0) \hat F_{jk}^{N}(n_0)\right]
\end{equation}
where the second term in RHS correspond to $a_0a^3F_{10}F_{20}$. More generally, operators for $F_{i0}F_{j0}$ are
\begin{equation}
    \Tr\left[\hat F^N_{i0}\hat F^N_{j0}(n_0)\right] = \frac{g_s^2}{a^4}\Tr[\hat \pi_{n_0,i}\hat \pi_{n_0,j}]
\end{equation}
Thus the naive $\hat T_{ij}(n_0)$ in the Hamiltonian formulation is
\begin{equation}
    \hat T^N_{ij}(n_0) = -\frac{g_s^2}{a^4} \Tr\left[\hat \pi_{n_0,i}\hat \pi_{n_0,j}\right] + \Tr\left[\hat F^{N}_{ik}(n_0)\hat F^{N}_{jk}(n_0)\right]\text.
\end{equation}

\begin{figure}[b]
    \centering\includegraphics[width=0.55\linewidth]{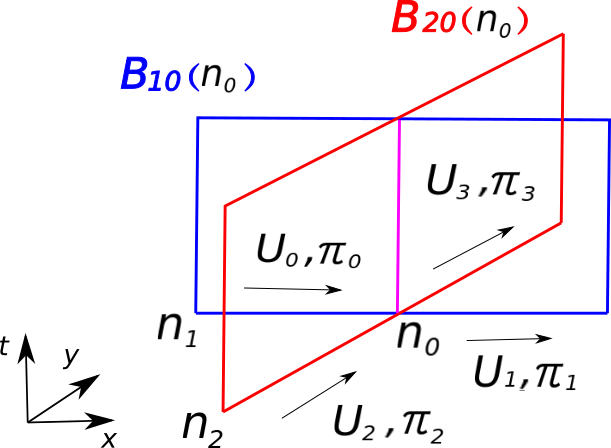}
    \caption{Half-clovers $B_{10}(n_0)$ and $B_{20}(n_0)$ at site $n_0$.} 
    \label{links1} 
\end{figure}

The clover approximations are obtained from $F_{ij}^{C}$ in \eq{fc} and $F_{i0}^{B}$ in \eq{fb}. As before, the transition from the action formalism to the Hamiltonian is straightforward for $F_{ij}$, so we focus only on the $F_{10}F_{20}$ term. For these, 
\begin{eqnarray}
    K_{\epsilon}(U',U) &=& K+ \epsilon a_0a^3  \Tr\left[F^{B}_{10}(n_0)F^{B}_{20}(n_0)\right].
\end{eqnarray}
We use the definitions of Fig.~\ref{links1} for the links around $n_0$. For the example of $U_{n_0,1}$, we denote operators and functions on them as $U_1, \hat U_1, \hat \pi_1$, and $g_1 = e^{ix_1^a\lambda^2}$.
Then $\hat T_\epsilon$ is 
\begin{widetext}
\begin{equation}
\hat T_\epsilon = \int\mathcal{D}g
\, e^{i K(g)-i\frac{\epsilon a}{16g_s^2a_0}\text{Tr}\left[ (g_{1}^{\dagger}-g_{1}+\hat U_0^{\dagger}(g_0^{\dagger}-g_0)\hat U_0)(g_{3}^{\dagger}-g_{3}+\hat U_2^{\dagger}(g_2^{\dagger}-g_2)\hat U_2) \right] + i \hat V}
\end{equation}
\end{widetext}
After the saddle-point approximation around $x = 0$, $\hat T_\epsilon$ simplifies and  becomes
\begin{eqnarray}
\hat T_\epsilon&\sim& \int dx\; e^{i x_{\rho} \hat\pi_{\rho} - x_{\rho} M_{\rho\sigma}  x_{\sigma}+i\hat V}  = A e^{-\frac{1}{4}\hat\pi_{\rho}M^{-1}_{\rho\sigma}\hat\pi_{\sigma}+i\hat V} \nonumber\\
&& \mbox{\,\,with\,\,}M_{\rho\sigma} =- \frac{ia}{2g_s^2a_0}\delta_{\rho\sigma} - \frac{i\epsilon a}{4g_s^2a_0} (M_1)_{\rho\sigma}\label{eq:it3}
\end{eqnarray}
Matrix elements of $(M_1)_{\rho\sigma}$ are zero other than
\begin{equation}
\begin{aligned}
\left( M_1 \right)_{(n_0,1,a)(n_0,2,b)} &= \delta_{ab}\\
\left( M_1 \right)_{(n_0,1,a)(n_2,2,b)} &= \Tr\left[\lambda^a\hat U_2^{\dagger}\lambda^b\hat U_2 \right]\\
\left( M_1 \right)_{(n_1,1,a)(n_0,2,b)} &= \Tr\left[\hat U_0^{\dagger}\lambda^a\hat U_0\lambda^b \right]\\
\left( M_1 \right)_{(n_1,1,a)(n_2,2,b)} &= \Tr\left[\hat U_0^{\dagger}\lambda^a\hat U_0\hat U_2^{\dagger}\lambda^b\hat U_2 \right]
\end{aligned}
\end{equation}
where sites $n_1,n_2$ are as labeled in Fig.~\ref{links1}.
Now by expanding \eq{it3} to linear order in $\epsilon$, we obtain the perturbed Hamiltonian
\begin{align}
H_\epsilon =& H_{K,S} - \epsilon \frac{g_s^2}{4a}\bigg(   
\Tr\left[\hat\pi_{1}\hat\pi_{3}  \right]+  \Tr\left[\hat\pi_{1}\hat U_2^{\dagger}\hat \pi_{2}\hat U_2 \right] \nonumber\\
&+ \Tr\left[\hat U_0^{\dagger}\hat\pi_{0}U_0\hat\pi_{3} \right]+ \Tr\left[\hat U_0^{\dagger}\hat\pi_{0}\hat U_0\hat U_2^{\dagger}\hat\pi_{2}\hat U_2 \right]\bigg) .
\end{align}
Thus $F_{i0}F_{j0}$ is generally implemented as
\begin{align}
\Tr&\left[\hat F^B_{i0}\hat F^B_{j0}(n_0)\right] =  \frac{g_s^2}{4a^4}\bigg(  
\Tr\left[\hat\pi_{n_0,i}\hat\pi_{n_0,j}  \right]\nonumber\\
&+ \Tr\left[\hat\pi_{n_0,i}\hat U_{n_0 - \hat j,j}^{\dagger}\hat\pi_{n_0 - \hat j,j}\hat U_{n_0 - \hat j,j} \right]\nonumber\\
&+\Tr\left[\hat U_{n_0 - \hat i,i}^{\dagger}\hat\pi_{n_0 - \hat i}\hat U_{n_0 - \hat i,i}\hat\pi_{n_0,j} \right]\nonumber\\
&+\Tr\left[\hat U_{n_0 - \hat i,i}^{\dagger}\hat\pi_{n_0 - \hat i,i}\hat U_{n_0 - \hat i,i}\hat U_{n_0 - \hat j,j}^{\dagger}\hat\pi_{n_0 - \hat j,j}\hat U_{n_0 - \hat j,j} \right] \bigg).\label{eq:f10fj0B}
\end{align}
The spatial $F_{ik}F_{ik}$ via clovers can be directly converted to operators.
With these operators, operators for measuring the spatial off-diagonal components of $T_{\mu\nu}$ are fully constructed following \eq{tij}.\\

\subsection{\texorpdfstring{$T_{0i}$}{T0i} in the Hamiltonian formulation}\label{sec:T0i}
In this subsection we derive Hermitian operators for $T_{0i}$, whose all terms contain time-like plaquettes and thus need to be appropriately converted for quantum simulations. As an example, $T_{01}$, via naive discretization, is written as:
\begin{eqnarray}
T_{01} &=& \Tr\left[F_{02}F_{12} + F_{03}F_{13}\right] \\
&=&  \Tr\left[F^{N}_{02}F^{N}_{12} + F^{N}_{03}F^{N}_{13}\right]  + O(a)
\end{eqnarray}
To find operators for $T_{01}$, As two terms are in the same form $F^{N}_{0j}F^{N}_{ij}$, let us perturb the Wilson action with only $F^{N}_{02}F^{N}_{12}$ at particular site $n_0$ to obtain
\begin{eqnarray}
    K_{\epsilon}(U',U) &=& K+ \epsilon a^3 a_0 \Tr\left[F^{N}_{02}F^{N}_{12}\right]
\end{eqnarray}
from which we find the following $\hat T_\epsilon$:
\begin{equation}
\hat T_\epsilon = \int\mathcal{D}g
\,\, e^{i K(g)-\frac{i\epsilon}{4g_s^2}\text{Tr}\left[ (g_{n_0,2}-g_{n_0,2}^{\dagger})(\hat P_{12}(n_0) - \hat P_{12}^{\dagger}(n_0)) \right] + i\hat V}.
\end{equation}
In the limit of $a_0 \rightarrow 0$, one can approximate the integral via saddle-point approximation around $x=0$:
\begin{eqnarray}
\hat T_\epsilon&\sim& \int dx\;  e^{\frac{ia}{2g_s^2a_0}x_{\rho} \delta_{\rho\sigma}  x_{\sigma} + i x_{\rho} \hat\pi'_{\rho} + i\hat V} \\
\text{where }\hat \pi'_{\rho} &=&  \hat\pi_{\rho} + \frac{\epsilon}{g_s^2}\delta_{nn_0}\delta_{i2}\Tr\left[ \lambda^b\im \hat P_{12}(n_0) \right]\text. \nonumber
\end{eqnarray}
Evaluating the Gaussian integral gives, at $O(\epsilon)$, 
\begin{equation}
\hat H_\epsilon = \hat H_{KS} + \epsilon \frac{1}{a}\Tr\left[ \hat \pi_{n_0,2} \im \hat P_{12}(n_0) \right] \text.
\end{equation}
From the perturbed Hamiltonian, we read off
\begin{equation}
   \Tr\left[\hat F^N_{02}\hat F^N_{12}(n_0)\right] = - \frac{1}{a^4}\Tr\left[ \hat \pi_{n_0,2} \im \hat P_{12}(n_0) \right]\text.
\end{equation}
Therefore, the operator to give $T_{0i}$ with naive discretization is 
\begin{equation}
    \hat T_{0i}(n_0) =  -\sum_{j\neq i}\frac{1}{a^4}\Tr\left[ \hat \pi_{n_0,j} \im \hat P_{ij}(n_0) \right]\text.
\end{equation}

\begin{figure}[b]
\centering
    \includegraphics[width=0.6\linewidth]{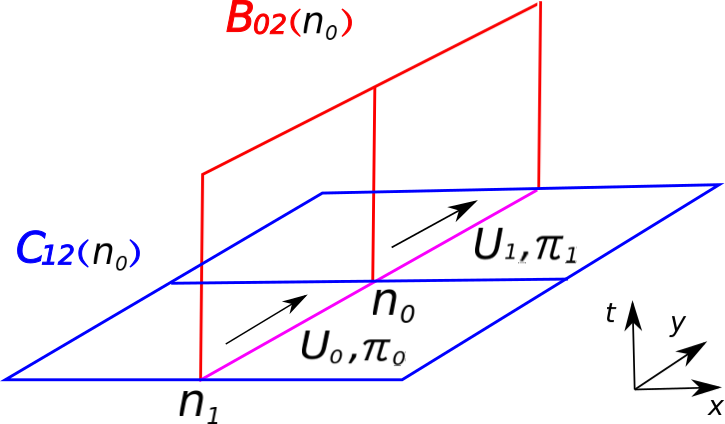}
    \caption{Clover $C_{12}(n_0)$ and half-clover $B_{02}(n_0)$ at site $n_0$.} 
    \label{links2}
\end{figure}

The naive discretization induces $O(a)$ error in $T_{0i}$. To improve up to $O(a^2)$, we use the clovers, $F_{ij}^{C}$ in \eq{fc} and $F_{i0}^{B}$ in \eq{fb} instead. Again as an example, we add $F_{02}F_{12}$ to the Wilson action to derive the transfer matrix from a perturbation given via
\begin{eqnarray}
    K_{\epsilon}(U',U) &=& K+ \epsilon a^3 a_0 \Tr\left[F^{B}_{02}(n_0)F^{C}_{12}(n_0)\right]
\end{eqnarray}
In the following we introduce notation for links around the site $n_0$ as in Fig.\ref{links2}.
The transfer matrix is then
\begin{widetext}
\begin{equation}
\hat T_\epsilon = \int\mathcal{D}g\; e^{i K(g)-i\frac{\epsilon }{8g_s^2}\text{Tr}\left[ (g_{1}-g_{1}^{\dagger}+\hat U_0^{\dagger}(g_0-g_0^{\dagger})\hat U_0)(\hat C_{12}(n_0)-\hat C_{12}^{\dagger}(n_0)) \right] + i \hat V}
\end{equation}
\end{widetext}
where $\hat C$ is the clover operator. In the limit of $a_0 \rightarrow 0$, one can evaluate the integral via the saddle point around $x=0$:
\begin{eqnarray}
\hat T_\epsilon\sim \int &dx&\; e^{\frac{ia}{2g^2a_0}x_{\rho} \delta_{\rho\sigma}  x_{\sigma} + i x_{\rho} \hat \pi'_{\rho} + i\hat V} \\
\hat \pi'_{\rho} =  \hat \pi_{\rho} &+& \frac{\epsilon}{2g_s^2}\delta_{nn_0}\delta_{i2}\left(\Tr\left[ \lambda^a\im \hat C_{12}(n_0) \right] \right.\nonumber\\
&+& \left. \Tr\left[ \hat U_0^{\dagger}\lambda^a \hat U_0\im \hat C_{12}(n_0) \right] \right)
\end{eqnarray}
which yields to $O(\epsilon)$, 
\begin{eqnarray}
\hat H_\epsilon =  \hat H_{KS} &+& \epsilon \frac{1}{2a}\left(\Tr\left[ \hat \pi_{1} \im \hat C_{12}(n_0) \right]\right.\nonumber\\
&+&\left.\Tr\left[ \hat U_0^{\dagger}\hat \pi_{0} \hat U_0\im \hat C_{12}(n_0) \right] \right)
\end{eqnarray}
From $\hat H_\epsilon$, we read off, for general $F_{0j}F_{ij}(n_0)$,
\begin{eqnarray}
    \Tr\left[\hat F^B_{0j}\hat F^C_{ij}(n_0)\right] = -\frac{1}{2a^4}\left( \Tr\left[ \hat \pi_{n_0,j} \im \hat C_{ij}(n_0) \right]\right.\nonumber\\
    +\left. \Tr\left[ \hat U_{n_0-\hat j,j}^{\dagger}\hat \pi_{n_0-\hat j,j} \hat U_{n_0-\hat j,j}\im \hat C_{ij}(n_0) \right]\right)\text.
\end{eqnarray}
From this operator, $T_{0i}(n)$ can be fully constructed.

\section{QGP state preparation}\label{sec:stateprep}
We have described the operators, acting on the Hamiltonian lattice, that provide the shear viscosity transport coefficients. In order to evaluate these expectation values, we need circuits corresponding to time evolution under the physical Hamiltonian; such circuits are described for a general gauge theory in~\cite{Lamm:2019bik}.  In addition we need to prepare states for them to act on which will enable one to reproduce the appropriate thermal expectation values.  The purpose of this section is to describe several practical methods for this thermal state preparation.

Before describing the methods, it is worth examining whether there are \emph{in principle} difficulties preparing a thermal state, such that the preparation necessarily requires time exponential in the volume. This is famously the case for certain frustrated spin systems~\cite{troyer2005computational} at low temperatures. If a similar barrier exists for thermal state preparation in gauge theories, our methods will be of no practical use.

To see there is unlikely to be a problem with the preparation of a QCD (or Yang-Mills) thermal state, it is useful to appeal to experiment.  A key feature of heavy-ion phenomenology is that after the initial hard interactions, the system rapidly equilibrates into a system that is well-described by hydrodynamics.  The equilibration time is believed to be $\sim 1\;\mathrm{fm}$; comparable to the characteristic scale of QCD. It is highly plausible that such rapid equilibration is a generic feature of strongly-coupled theories~\cite{PhysRevLett.124.102301,Dore:2020jye}.  Accordingly, one expects that if an out-of-equilibrium state is prepared, it equilibrates on roughly the natural time scales of the theory.  Thus, it seems implausible that preparing a thermal state would take an exponentially long time.

It is particularly convenient that the thermal states of interest to us have $T\sim 200\;\mathrm{MeV}$. Very cold temperatures, near the true ground state, may be difficult to prepare due to frustration. Very high temperature states ($T\gg \Lambda_{\mathrm{QCD}}$) likely have difficulty thermalizing, as the fluid is closer to a free gas. At $T\sim 200\;\mathrm{MeV}$, the strongly interacting fluid is at a temperature comparable to other physical scales. Even without experimental evidence, we might expect such systems to equilibrate quickly.

Given this discussion it is natural to assume that polynomial-time thermal state preparation is possible --- even lacking a formal proof. With this assumption, we now turn to the task of finding a practical way to construct such a state.

In this section several approaches are explored to prepare thermal states.  We explore multiple paths since at this stage the relative merits with respect to future quantum devices is unknown. One general note applies to all methods discussed here. Common digitization schemes result in many possible states on the quantum computer that correspond to no physical state. An important example is gauge invariance: the physical states are those that are unchanged by any gauge transformation. Naively, this is a serious difficulty, as we need to ensure that the prepared state is one of the rare gauge-invariant ones. However, as discussed in~\cite{Lamm:2019bik,Halimeh:2020ecg,Tran:2020azk} and elsewhere, because time-evolution (even under a Suzuki-Trotter approximation) commutes with gauge transformations, any state preparation method that builds a gauge-invariant state and then performs time-evolution will automatically respect the gauge symmetry up to quantum noise.

\subsection{Thermal states}

Before looking at detailed schemes, it behooves us to first consider what we mean by a thermal state. Ideally, we want a single quantum state in which expectation values match those given by the canonical ensemble $\langle \mathcal O \rangle = Z^{-1} \Tr \;e^{-\beta H} \mathcal O$.
In such a state the physical system --- in this case of the appropriately truncated version of the lattice gauge theory --- is modeled by a subspace of a larger system. We can define a thermal state as
\begin{equation}
        |T\rangle = \sum_k c_k |\psi_k\rangle_{\rm sys} \otimes |\psi_k\rangle_{\rm comp}
\end{equation}
where a subscript of $\mathrm{sys}$ denotes a state of the physical system being studied, and one of $\mathrm{comp}$ denotes the state of the complementary system. The coefficients $c_k$ are fixed by considering what happens when we trace out the complementary system; in particular, we require
\begin{equation}
    \rho_T^{\rm sys}\equiv{\rm Tr^{\rm comp}}\left [ |T\rangle \langle T| \right ] = \frac{ \sum_k e^{-\beta E_k} |\psi_k\rangle_{\rm sys} \, {}_{\rm sys}\langle\psi_k|}{\sum_j e^{-\beta E_j} }  
\label{densityM}\end{equation}
so that for an operator ${\cal O}^{\rm sys}$ that acts entirely in the system subspace, the quantum expectation values match the desired thermal ones. Thus, the density matrix appears as the classically uncertain state of a quantum system once the complementary system has been traced out.

Obtaining a true thermal state is difficult, but generally unnecessary: a good approximation is sufficient. In fact, there are a wide variety of possible ensembles, all of which agree in the thermodynamic limit --- see~\cite{Sugiura:2013pla} for an elegant example. By exploiting the equivalence between microcanonical and canonical descriptions\footnote{Here, we are not considering other conserved quantities in the system so that the grand canonical ensemble is not relevant.} in the thermodynamic limit, we should be able to learn about systems even when we cannot obtain a distribution of energies with probabilities given by Boltzmann weights.

In the microcanonical approach to statistical mechanics, one computes properties at a fixed energy rather than at a fixed temperature. Given the standard assumptions relating statistical mechanics to thermodynamics, in the thermodynamic limit of large systems, quantities computed in the microcanonical and canonical descriptions will agree~\cite{pathria2011statistical}. For `typical' quantum field theories (presumably including both Yang-Mills and QCD), given an eigenstate $|\Psi\rangle$ of the system Hamiltonian $H^{\mathrm{sys}}$ sampled from the set of all eigenstates with energy density is near $\epsilon$, expectation values $\langle\Psi|\mathcal O|\Psi\rangle$ will agree in the thermodynamic limit with the canonical values.

This suggests a cheap approach to obtain thermodynamic expectation values. Rather than preparing a mixed state that exactly reproduces the canonical ensemble, we prepare a typical pure state with the desired energy density. This can be done, in practice, by preparing \emph{any} pure state with the desired energy density, and then time evolving to allow the state to thermalize.

\subsection{The heat bath approach}

While there is no practical way of constructing a perfect thermal state $|T\rangle$, it is quite straightforward to find an algorithm to create a reasonable approximation.  The method exploits the physical principle of a heat bath.  The key idea is that one considers a very large total system --- much larger than the system subspace --- with the complementary subspace serving as heat bath. The heat bath is subject to dynamics under the Hamiltonian $H^{\rm HB}$, which allows explicit construction of its eigenstates.  

One starts with initial conditions in which the heat bath and the system are decoupled; the heat bath is prepared in a known initial state of low energy and the system is in some high-energy state.  This physical state is chosen primarily for simplicity --- the details of the state will not matter provided that the heat bath is sufficiently large. The heat bath and the system of interest are dynamically coupled in a gauge invariant way via a Hamiltonian, $H^{\rm couple}$, which is initially switched off. The coupling Hamiltonian satisfies
\begin{equation}
    [H^{\rm couple}, H^{\rm sys}] \ne 0 \;, \; \;  [H^{\rm couple}, H^{\rm HB}] \ne 0\text,
\end{equation}
so that when it is switched on the coupling allows energy to flow between the system and the heat bath.  If one waits sufficiently long, one might reasonably expect the system to approximately thermalize.
The $H^{\rm couple}$ must be small in the sense that at all stages in the evolution the absolute value of its expectation value: 
\begin{equation}
   \langle H^{\rm couple}\rangle \ll \langle H^{\rm sys} \rangle -     \langle H^{\rm sys} \rangle_{\rm vac} \; . \label{CoupleScale}
\end{equation}
where $\langle H^{\rm sys} \rangle_{\rm vac}$ is the expecation value of the vacuum state. This condition ensures that the details of the coupling between the heat bath and the system has a negligible effect on the final result.

Once the system has thermalized,  one can switch off $H^{\rm couple}$.   This is essentially modeling the physical process by which physical systems thermalize; and the basic assumption underlying statistical mechanics is that the details of how the system thermalizes should not matter.  By choosing various initial configurations of the heat bath,  one can evolve the system into (approximately) thermalized systems at various temperatures.

While this method should work as a matter of principle, it has a strong practical disadvantage: it requires a  very large number of qubits to encode the heat bath.  Of course, the general notion of a heat bath in thermodynamics is that it should be large.  Moreover, in the present context there are some particular challenges requiring this.

Recall that $|T\rangle$ formally requires a heat bath at least as large as the system; otherwise the condition $ {}_{\rm HB} \langle  \psi_k'|\psi_k\rangle_{\rm HB}=\delta_{k',k}$ cannot be met.  However as a practical matter the heat bath is coupled to the system in an passive way; it does not couple system states to heat bath states in a manner that naturally pushes the system towards configurations satisfying  ${}_{\rm HB} \langle  \psi_k'|\psi_k\rangle_{\rm HB}=\delta_{k',k}$.  Rather to achieve some approximation to that condition, one requires a large heat bath subspace in which approximate orthogonality is likely to emerge naturally.  

There is a second practical issue that suggests the need for large heat baths.  For a generic interacting theory, the only states we know how to write explicitly in a practical way are states with energies at the scale of the lattice spacing.  Thus, to cool the system to a temperatures of physical interest one needs to transfer a substantial energy from the system to the heat bath.  This in turn means that the heat capacity of the heat bath must be large.

While heat bath must be large, it is clear that its size scales polynomially in the physical size and lattice spacing of the system.  Thus, the heat bath is a useful demonstration of an explicit method with polynomial scaling.  However, given the large size of heat bath that the approach requires, it is unlikely to be the optimal way to pursue thermal physics on quantum computers.

\subsection{Active cooling}

A principal problem with the heat bath approach is that it is essentially passive.  The only active step is the connecting the system with the heat bath through $H^{\rm couple}$.  One reason the heat bath needs to be large in such a passive scheme is that it must absorb all of the excess energy of the system, which will start at an energy density of lattice scales. 

Let us consider a slight variation on the heat bath approach. Suppose that instead of a single large heat bath one had $N$ smaller heat baths each starting in its ground state configuration.  Moreover, let the system begin in a state with $\langle H^{\rm sys}\rangle=E_0$.  We connect the system to the first heat bath alone, via a small coupling Hamiltonian. The coupled system evolves dynamically for some time after which this coupling is switched off.  Assuming that the coupling Hamiltonian is small in sense of Eq.~(\ref{CoupleScale}), the dynamics must reduce the energy of the system: the energy of the coupled system is conserved and the energy of the complementary system can only increase since it starts in the ground state.  Thus at the end of the time which the first heat bath is coupled to the system $\langle H^{\rm sys}\rangle=E_1$  with $E_1 < E_0$. 

At that stage, the system is decoupled from the first heat bath and coupled to the second. We perform more time evolution such that $\langle H^{\rm sys}\rangle = E_2$, with $E_2<E_1$.  Continuing the process, we see that the energy is monotonically decreasing:
\begin{equation}
    E_N < E_{N-1}  \cdots < E_2 < E_1 < E_0 \; .
\end{equation}
Including enough smaller heat bath allows the system to lose as much energy as one wishes provided the coupling Hamiltonian is sufficiently small.

Superficially, this approach --- like the heat bath approach --- is essentially passive.  The only active step appears to be the coupling and decoupling of the system with the various heat baths.  This is misleading, however, there was another active step:  preparing the various heat baths to be in their ground state.  If one was not able to do that, the approach would not be viable.  However, if one can initialize these into their ground states, there is no need to use multiple different ``heat baths''; rather the same ``heat bath'' can be reused and reinitialized to its ground state between cycles.  The system will yield precisely the same the results as it would have had one used multiple ``heat bath''.   The virtue of this is that qubit costs needed obtain a typical energy density for the system is greatly reduced compared to the heat bath method.

In what follows, we refer not to a ``heat bath'' but suggestively to a ``pump'' since it does not keep the system in thermal equilibrium.  The pump can be of similar size to the physical system --- or smaller --- and the key point is that it absorbs energy from the system, moves the energy into the environment and is reinitialized to its ground state.  This step of reintitalizing the complementary system is an active one. In effect the pumps is acting as quantum heat pump or refrigerator~\cite{rossnagel2016single} (for a review of quantum refrigerators see \cite{frig}), thus the name. 

Of course, there is no way for such a device to act entirely via unitary evolution within the Hilbert space of qubits.  The act of the pump dropping back into its ground state is clearly not unitary.  It is worth recalling that quantum computing requires more than just control of the unitary behavior of qubits: one of the DiVincenzo criteria~\cite{divincrit} is the ability to initialize the qubits to some fiducial state. This initialization is not unitary and necessarily involves entangling the states of the Hilbert space of the quantum computer with the environment. This method requires that the one can separately initialize the system and the pump, and that the initialization (which as we will see requires measuring the pump qubits) of the pump can be done much faster than the physical system can decohere -- a nontrival device specification.

It is useful to have a concrete model in mind for the purposes of modeling.  Conceptually, if not practically, the simplest way to think of this is to choose the fiducial state of the pump as $|g\rangle^{\rm pump} =|0,0,0, \cdots 0,0, 0\rangle$ and then to choose $H^{\rm pump}$ so its ground state is the fiducual state.  One can reinitialize by simply measuring the qubits  using $\sigma_z$ to determine if the qubit is in $|1\rangle$ or $|0\rangle$  and if the result is $|1\rangle$ act on it with the unitary operator $\sigma_x$.  In effect, this can be viewed as a qubit version of Maxwell's demon, and a cycle that employed such a device might be referred to as a quantum demon refrigerator.

Let us provide a simple model of a cycle which will lower the system energy which started from a state with a lattice-scale energy density. The Hilbert space is spanned by outer product states of a space characterizing the system and the pump, which we will now assume to be comparable in size.  Again we have three Hamiltonian, $H^{\rm sys}$ that acts on the system, $H^{\rm pump}$, that acts in the pump space and $H^{\rm couple}$ that couples the two.  The coupling dynamics is constructed to be gauge invariant, $H^{\rm couple}$ must have non-zero commutators with $H^{\rm sys}$ and $H^{\rm pump}$
and must be small in the sense of satisfying inequality (\ref{CoupleScale}).

\begin{figure}
    \centering
\vspace{0.5cm}
\includegraphics[width=\linewidth]{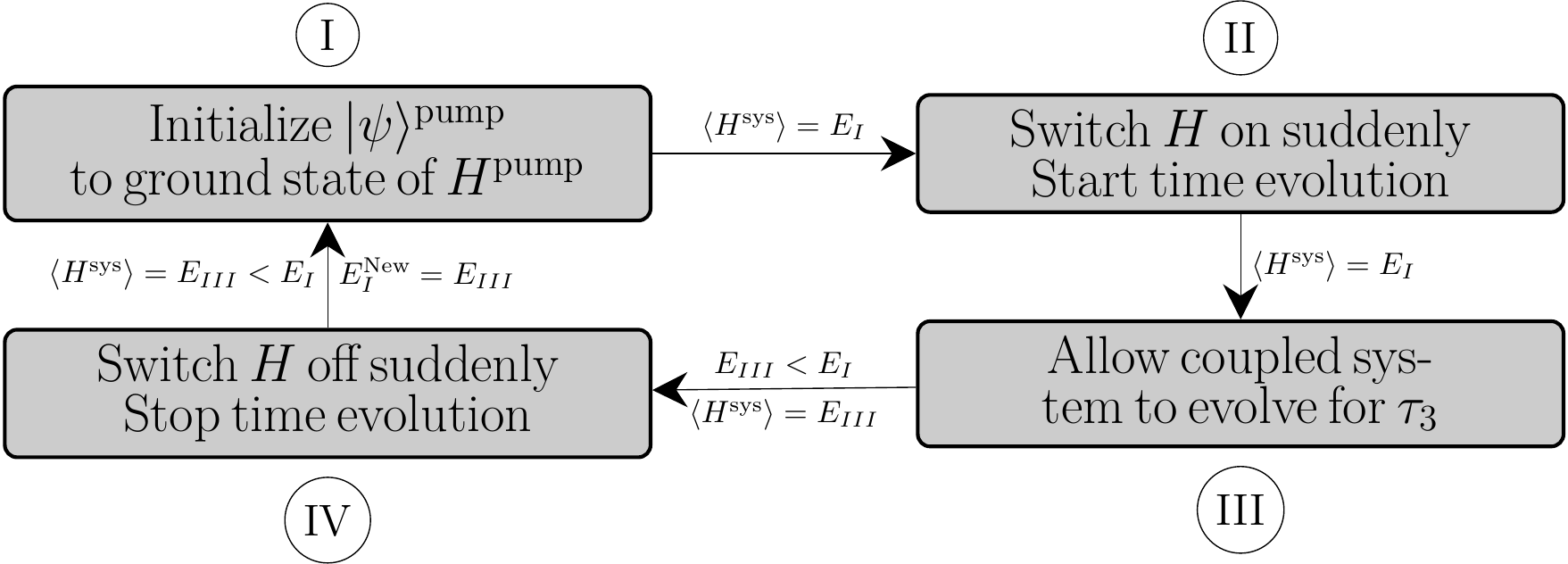}
    \caption{An active cooling cycle. \label{fig:cycle}}
\end{figure}

A schematic description of an active cooling cycle is given in Fig.~\ref{fig:cycle}.  In step I the time evolution has been switched off, the system begins in some pure or mixed state with $\langle H^{\rm sys} \rangle =E_I$ relative to its ground state, the pump subspace is then initialized to its ground state,  with $\langle H^{\rm pump} \rangle =0$.     Since $H^{\rm couple}$  satisfies inequality (\ref{CoupleScale}),  the total energy (relative to the ground state) of the combined system  $\langle H^{\rm sys} + H^{\rm pump} +H^{\rm couple} \rangle$ , is also $E_I$ up to a small correction.  In step II, the time evolution associated with the combined Hamiltonian is switched on suddenly, since this is sudden $\langle H^{\rm sys} + H^{\rm pump} +H^{\rm couple} \rangle$ remains at $E_I$.  In step III the combined system undergoes time evolution under $H^{\rm sys}+H^{\rm pump}+H^{\rm couple}$ for a fixed time.  During this time evolution, the total energy is conserved and any net energy flow goes from the system to the pump. After some time, $ \langle H^{sys}\rangle $ reaches some value $E_{III}$ where $E_{III} < E_I$.   The time evolution then is switched off suddenly leaving $ \langle H^{sys}\rangle =E_{III} < E_I$.  At that point, the cycle repeats.  Thus each time through the cycle, $ \langle H^{sys}\rangle$ drops.  One could continue iterating the cycle and lowering $\langle H^{sys}\rangle$ until the inequality (\ref{CoupleScale}) ceases to hold.

Consider the reduced density matrix of Eq.~(\ref{densityM}) for the system as after tracing over the pump. By construction step I preserves $\rho^{\rm sys}_T$. This is obvious since during there is no coupling between the system and the pump.  

Suppose at step I the full density matrix, $\hat{\rho}^I$ is given by 
\begin{equation}
    \hat{\rho}^I = \sum_{a,a',b,b'}  |a'\rangle^{\rm sys} |b'\rangle^{\rm pump} \, \hat{\rho}_{a,b;a'b'} \,  {}^{\rm pump}\langle b | {}^{\rm sys} \langle a|
\end{equation}
where the states $|a\rangle^{\rm sys}$ and $|b\rangle^{\rm pump}$ are states in an orthonormal basis for the system space and complementary space respectively.  Thus matrix elements of the reduced matrix elements are given by 
\begin{equation}
 \hat{\rho}^{I \,\rm sys}_{a,a'} = \sum_{b,b',b''}     {}^{\rm pump}\langle b'' ||b \rangle^{\rm pump} \hat{\rho}_{a,b;a'b'}  {}^{\rm pump}\langle b'' ||b \rangle^{\rm pump}
\end{equation}

Thus the full density matrix at step II is given by 
\begin{widetext}
\begin{equation}
 \hat{\rho}^{II} =  \sum_{a,a'}   |a'\rangle^{\rm sys} |g\rangle^{\rm pump} \,\left( \sum_{b,b',b''}     {}^{\rm pump}\langle b'' ||b \rangle^{\rm pump} \hat{\rho}_{a,b;a'b'}^{I}  {}^{\rm pump}\langle b'' ||b \rangle^{\rm pump} \right ) {}^{\rm pump}\langle g| {}^{\rm sys} \langle a|
\end{equation}
\end{widetext}
where $|g\rangle^{\rm pump}$ is the ground state of the pump.  In the remainder of the cycle $\hat{\rho}$  evolves to $\hat{\rho}^{III} = U_{III} ^\dagger \rho_{II}  U_{III}$ with $U_{III} =\exp \left ({-i(H^{\rm sys} + H^{\rm pump} +H^{\rm couple} ) \tau_3}\right )$
where $\tau_3$ is the time the system evolves for in step III.

In the active cycle, energy is pumped out of the combined system (system plus pump) and into the environment by the act of initializing the pump. In the process the entropy of the pump drops and thus the entropy of the environment increases. For that reason we label this approach an active cooling cycle. One might quibble that this is a bit of a misnomer since the system is not thermally equilibrated and thus, it is not clear that the energy pumped out can be accurately described as heat.   But the essence of this active cycle is very much the same as the cooling in a quantum refrigerator. Moreover as described below one can use a variation on this approach to achieve a good approximation to a true thermal equilibrium.

Clearly, this active cycle approach is rather general and variations on this theme can be developed. As formulated, the approach requires explicit choices for the size of the pump as well as for the form and strength of $H^{\rm couple}$, and $H^{\rm pump}$, and $\tau_3$.   It is clear that to get high performance with this method one must choose these well. It is an open question as to what optimal choices are for these.

One obvious approach is to tailor the overall strength of $H^{\rm couple}$ to the iteration. There is a trade off: strong coupling leads to rapid transmission of energy from the system to the pump and reduce the overall time for reducing the energy.  However, this comes at cost; large coupling limits the lowest energy density of the system one can achieve.  The cycle can only be shown to remove energy from the system when the energy in $H^{\rm couple}$ is negligible.  Thus a sensible approach would be to make $H^{\rm couple}$ large during early cycles in order to facilitate rapid energy transfer and in later cycles reduce it to allow reduction to lower energy densities.

A similar approach might be taken with regard to the pump.  There is a freedom to set the overall energy scale of $H^{\rm pump}$.   It is straightforward to see that $N^{\rm cycles}$, the number of cycles needed to go from $\langle H^{\rm sys} \rangle =E_{i}$ (presumably with energy density at the lattice scale) to a final configuration with $\langle H^{\rm sys} \rangle =E_{ f}$ scales logarithmically with the ratio of $E_i$ to $E_f$:
\begin{equation}
N^{\rm cycles} = {\rm A} \log \left ( \frac{E_i}{E_f} \right )
\label{Eq:Ncycle}\end{equation}
where $A$ is a numerical coefficient of order unity, provided that one tailors the value of the overall strength of the cycle to the cycle in an appropriate way.  

To see how this comes about, consider the trade offs involved in setting the scale of $H^{\rm pump}$. If it is set too large then it is difficult to induce transition in the pump and this energy flow will be very slow.  On the other hand if it is too small the maximum amount of energy that can be absorbed in a cycle is limited.  This is clearly true since the system is finite.  Moreover, at some point the energy flow from the system of interest to the pump becomes negligibly small (either because the system and pump are equilibrating towards zero net flow).  The amount of energy transferred before the energy flow becomes negligible will clearly depends sensitively on $H^{\rm pump}$. 

A simple compromise would be to choose the overall strength of $H^{\rm comp}$ to be large enough so that some modest fixed fraction, $f$ of the system energy at the beginning of the cycle is transferred before the energy flow becomes negligible. 
However, the exact value depends on the initial configuration of system with the strength increasing with $\langle H^{\rm sys} \rangle$.  Thus one might change the strength of each cycle to keep $f$ approximately the same in each cycle.   It would be natural to end each cycle well before the fraction of the energy, $f$, is transferred, since to the extent the system equilbrates the energy transfer slows down as the fraction approaches $f$.  For simplicity, let us assume that each cycle stops when the fraction of energy transferred is $f_c\times f$ with $1>f_c>0$.  One might, for example, take $f_c=\frac{1}{2}$.  Thus each cycle reduces by a factor of $1-f_c f$ and the number of cycles it takes to go from a configuration with  $H^{\rm sys}=E_{\rm i}$ to one with $\langle H^{\rm sys}\rangle =E_{\rm f}$ is thus given by 
Eq.~(\ref{Eq:Ncycle}) with
\begin{equation}
    A= \frac{1}{\log(1-f f_c)} \; .
\end{equation}

Of course, the  method outlined above is undoubted not the optimal way to reduce the energy of system of interest from $E_i$ to $E_f$ given various resource constraints.  But it clearly demonstrates that the minimum number of cycles needed can be quite modest since the optimal choice will scale no worse than logarithmically in $E_i/E_f$.

As given so far this approach can produce density matrices with $\langle H^{\rm sys}\rangle$ at energy densities of interest and this should allow for a microcanonical extraction of transport coefficients.  While this is sufficient for our purposes it is useful to note that  a variation on this method is likely to produce density matrices $\rho^{\rm sys}$ which, to good approximation, are thermal.  

The basic approach is to start from preceding approach and produce $\rho^{\rm sys}$ with  $\langle H^{\rm sys}\rangle$ that is at the correct general scale for the energy of the thermal ensemble at temperature $T$ that we wish to study.  This can be done in comparatively few cycles.

Next one continues to cycle but instead of reinitializing the pump to its ground state, each time reinitialize the pump to some well-defined state with= $E^{\rm pump}$.  This can be done by initializing to the ground state and then making unitary transformations to bring the pump to that state. By hypothesis, the dynamics of the pump is simple enough to do so.   The value of $E_{\rm pump}$ to be match the expectation value of $H^{\rm pump}$ in thermal equilibrium with a heat bath at temperature $T$.
\begin{equation}
\begin{split}
E^{\rm pump} & = - T^2\frac{ Z'(T)}{Z(T)} \; \; {\rm with}\\
Z(T) & = \sum_i e^{-i E_i^{\rm}/T} 
\end{split}
\end{equation}
If in each cycle, step III is allowed to last long enough for the system and the pump to equilibrate, then after multiple cycles one would expect that $\rho^{\rm sys}$ to become approximately thermal with temperature $T$.

\subsection{Adiabatic state preparation}

An entirely different approach to preparing a low-temperature thermal state begins from the adiabatic theorem~\cite{messiah1962quantum}. Adiabatic state preparation is a method for obtaining the \emph{ground} state of a Hamiltonian. Given the ground state of the physical Hamiltonian, it is an easy matter to add energy. Allowing the resulting system to thermalize, we can obtain a thermal state at any desired energy density.  A great benefit to adiabatic methods is that they require no ancillary qubits to perform the preparation. The primary difficulty is in obtaining the ground state.

Adiabatic state preparation begins with a Hamiltonian whose ground state is known, and can be prepared directly on the quantum computer.  One allows quantum time-evolution to progress while slowly changing the Hamiltonian from the initial one to the Hamiltonian of interest.  In principle, such a scheme is guaranteed to yield the ground state of the system of interest to good approximation provided that the Hamiltonian is varied slowly.  For gauge theories there is an additional constraint.  One wishes to find the ground state in the physical subspace.  As a result, it is natural to restrict all Hamiltonians in the path from the initial one to the final one to be invariant under spatial gauge transformations.

There is a generic, practical issue for any approach of this sort:  it requires a slow evolution and hence long times. These constraints are polynomial in all parameters, but nevertheless may result in significant hurdles for the forseeable future. Quantifying the time evolution needed is model-dependent and remains for future work.

An obvious state to start with is the weak-coupling limit of the lattice gauge theory. In the case of pure Yang-Mills theory, the ground state is the gauge-invariant projection of the ground state of $N^2-1$ free vector fields. The preparation of this state is described in~\cite{Lawrence:2020irw}. However, several difficulties may prevent this limit from being practical. In the weak coupling limit, the spatial volume being simulated is much smaller than the confinement scale. In order to reach the physical regime, this boundary must be crossed, and it is likely that a phase transition (associated with an exponentially small or even vanishing gap) is encountered. Even if there is no phase transition along the path to the desired coupling, the weak-coupling limit is inaccessible to many proposed truncation schemes for the gluon fields, including approximation by a finite subgroup~\cite{Alexandru:2019nsa,Ji:2020kjk} and momentum-space truncation~\cite{Ciavarella:2021nmj}.

The ground state of the strong-coupling limit is simpler to prepare~\cite{Lawrence:2020irw} and is natural in both the subgroup approximation and the character expansion. Furthermore, the strong-coupling limit is connected to realistic physical couplings without a phase transition in the infinite volume limit; therefore, the desired coupling can be reached without encountering a small gap, even one polynomially small in lattice units.

Other possible adiabatic paths may exist. One tempting possibility is to add a Higgs field, allowing a gap to be created in the weak coupling limit and avoiding the confinement transition.

\section{Discussion}\label{sec:discussion}

The simulation of heavy ion collisions requires large volumes $\sim 10\;\mathrm{fm}$ with fine resolutions  $\sim 0.1\;\mathrm{fm}$ evolved for long times $\sim 10\;\mathrm{fm}$ --- corresponding to enormous quantum resource requirements. In contrast, the transport coefficients can be extracted from the hydrodynamic limit of thermal states. Thus, such observables allow for simplified state preparation and fewer resources. In this work, we propose an algorithm for extracting the transport coefficients in lattice gauge theory from thermal states. In order to implement this, it was necessary to derive a lattice version of the stress-energy tensor in the Hamiltonian formalism. Further, this algorithm requires the preparation of a thermal state.  Here, multiple viable methods have been suggested. Future work should investigate the relative merits of each state preparation method in specific theories. Naive estimates would suggest that with $\sim 10^4$ qubits, one could produce a rough calculation of the viscosity of 3+1d pure glue $SU(3)$~\cite{Alexandru:2019nsa,Ji:2020kjk}. Such a calculation could account for all systematic errors from lattice computations. In the near-term, 2+1d $\mathbb{Z}_N$~\cite{Yamamoto:2020eqi,Gustafson:2020yfe} and $D_4$~\cite{Lamm:2019bik,Harmalkar:2020mpd} could be computed with a more modest $\sim 10^2$ and $\sim 10^3$ qubits respectively.

\begin{acknowledgements}
We are indebted to Guy Moore and Paul Romatschke for many insightful comments on hydrodynamics. T.D.C.~and Y.Y.~are supported by the U.S.~Department of Energy under Contract No.~DE-FG02-93ER-40762. Y.Y.~is additionally supported by the Jefferson Science Associates 2020-2021 graduate fellowship program. H.L.~is supported by a Department of Energy QuantiSED grant. Fermilab is operated by Fermi Research Alliance, LLC under contract number DE-AC02-07CH11359 with the United States Department of Energy. S.L.~is supported by the U.S.~Department of Energy under Contract No.~DE-SC0017905.
\end{acknowledgements}

\bibliography{viscosity}

\end{document}